\documentclass[aps,prd,twocolumn,superscriptaddress,preprintnumbers,showpacs]{revtex4-1}
\usepackage{amsmath,amsfonts,amssymb,graphicx,color,times,bbm,psfrag,dsfont,
slashed,bigints,bm}
\usepackage[colorlinks=true,citecolor=blue]{hyperref}
\usepackage[normalem]{ulem}

\usepackage[latin9]{inputenc}

\newcommand{\be}{\begin{equation}}
\newcommand{\ee}{\end{equation}}
\newcommand{\mnras}{{Mon.Not.Roy.Astron.Soc.}}

\newcommand{\bea}{\begin{eqnarray}}
\newcommand{\eea}{\end{eqnarray}}

\newcommand\actaa{\ref@jnl{AcA}}
\newcommand\caa{\ref@jnl{ChA\&A}}
\newcommand\cjaa{\ref@jnl{ChJA\&A}}
\newcommand\jcap{\ref@jnl{JCAP}}
\newcommand\na{\ref@jnl{NewA}}
\newcommand\nar{\ref@jnl{NewAR}}
\newcommand\pasa{\ref@jnl{PASA}}
\newcommand\rmxaa{\ref@jnl{RMxAA}}
\newcommand\aap{Astron.Astrophys.}
\newcommand\aapr{\ref@jnl{A\&A~Rv}}
\newcommand\aaps{\ref@jnl{A\&AS}}

\newcommand{\beq}{\begin{equation}}
\newcommand{\eeq}{\end{equation}}

\begin{document}

\title{A reliable description of the radial oscillations of compact stars}
\author{F. Di Clemente}
\email{francesco.diclemente.uni@gmail.com}
\affiliation{Dipartimento di Scienze Fisiche e Chimiche, Universit\`a dell'Aquila, via Vetoio,
I-67010 Coppito-L'Aquila, Italy.}
\affiliation{INFN, Laboratori Nazionali del Gran Sasso, Via G. Acitelli,
22, I-67100 Assergi (AQ), Italy.}
\author{M. Mannarelli}
\email[correspondence at: ]{massimo@lngs.infn.it}
\affiliation{INFN, Laboratori Nazionali del Gran Sasso, Via G. Acitelli,
22, I-67100 Assergi (AQ), Italy.}
\author{F. Tonelli}
\email{francesco.tonelli@lngs.infn.it}
\affiliation{Dipartimento di Scienze Fisiche e Chimiche, Universit\`a dell'Aquila, via Vetoio,
I-67010 Coppito-L'Aquila, Italy.}
\affiliation{INFN, Laboratori Nazionali del Gran Sasso, Via G. Acitelli,
22, I-67100 Assergi (AQ), Italy.}

\begin{abstract}
We develop a numerical algorithm for the solution of the Sturm-Liouville differential equation governing the stationary radial oscillations of nonrotating compact stars. Our method is based on the  Numerov's  method that  turns the Sturm-Liouville differential equation in an eigenvalue problem. In our development we  provide a strategy to correctly deal with the  star boundaries and  the interfaces between layers with different mechanical properties.  Assuming that the fluctuations obey the same equation of state of the background, we analyze various different stellar models and  we precisely determine hundreds of eigenfrequencies and of eigenmodes.    If the equation of state does not present an interface  discontinuity, the fundamental radial eigenmode becomes unstable exactly at the critical central energy density corresponding to the largest gravitational mass. However, in the presence of an interface  discontinuity, there exist stable  configurations with a  central density exceeding the critical one and with a smaller gravitational mass. 
\end{abstract}
\maketitle


\section{Introduction} 
The relativistic  equilibrium  of nonrotating stars can be  determined solving the Tolman-Oppenheimer-Volkoff (TOV) equation~\cite{Tolman, Oppenheimer-Volkoff} that expresses the  balance between the internal hydrostatic pressure and the gravitational pull. Actually, the TOV's equation provides  stationary hydrostatic solutions which may or may not correspond to stable configurations. Starting from a stellar configuration with a  small baryonic mass, by increasing the central matter density one obtains a sequence of TOV's stationary stellar solutions  with increasing gravitational  mass. However,  at a critical  central energy density  the gravitational mass reaches a maximum and then further increasing the central energy density the stellar mass starts to decrease because the gravitational binding energy dominates.  The configuration with the largest mass  is typically identified with the last stable configuration, indeed for larger values of the central energy density the general idea is that the  system should collapse~\cite{Shapiro:1983du}. 
The stellar collapse should be driven by growing radial oscillations: standing radial waves with an imaginary frequency, see~\cite{1983tsp..book.....C} for a general discussion. Therefore we find more appropriate to  define the last stable configuration as the one characterized by a null (or neutral) radial frequency. From the analysis of the spectrum of the radial oscillations one can determine whether the maximum mass configuration and the last stable configuration coincide. This is of a certain interest because  {\it twin} configurations, having the same gravitational mass but different radii, may exist.   

The equations governing the dynamical  stability  of the radial mode oscillations were  derived by Chandrasekhar in~\cite{Chandrasekhar:1964zza, Chandrasekhar:1964zz}  by a linear response  expansion. Then, they have been applied to various stellar models built using  different equation of states (EoSs), see for example~\cite{1965gtgc.book.....H, 1977ApJ...217..799C, 1983MNRAS.202..159G, 1992A&amp;A...260..250V, Gondek:1997fd,  Kokkotas:2000up, Camilo:2018goy}. In these analyses the typical  assumption is that  the radial oscillations are infinitesimal adiabatic perturbations of the stellar configurations, however the inclusion of nonlinearities  may qualitatively change the picture~\cite{1982AcA....32..147D, Gabler:2009yt}, with  a much richer dynamics and the possible stabilization of modes that are  linearly unstable. The unstable modes can also be damped by 
nonequilibrium processes,  see for instance~\cite{refId0}, which are particularly relevant for hot compact stars, see~\cite{Burgio:2011qe, Alford:2019qtm}, meaning that   in these cases the maximum mass and the last stable configuration do not coincide.

Since in the interior of compact stars the matter density is extremely high,  some exotic phases can be realized, including  meson condensation~\cite{Migdal:1971cu,Migdal_1972, Mannarelli:2019hgn} or quark deconfinement, see \cite{Rajagopal:2000wf,Alford:2007xm,Anglani:2013gfu} for reviews. Compacts stars composed of a  deconfined quark core surrounded by an envelope of nuclear matter are typically called   hybrid stars,  see for example~\cite{Glendenning:1997wn}.  Compact stars entirely composed of deconfined quark matter are instead called  strange stars~\cite{Alcock:1986hz, Haensel:1986qb}.  The analyses of the radial oscillations  of strange stars~\cite{1991MNRAS.250..679B, Datta:1992np, Gondek:1999ad,  VasquezFlores:2010zza, Salinas:2019fmu} and hybrid stars~\cite{1989A&amp;A...217..137H, Sahu:2001iv, VasquezFlores:2012vf, Brillante:2014lwa, Pereira:2017rmp} has shown that the typical oscillation frequencies are similar to those of standard neutron stars, reaching  the kHz range, although the high frequency mode  may be damped due to nonequilibrium weak process in the core~\cite{1989A&amp;A...217..137H}.  

The analysis of the radial oscillations of  hybrid stars,  or of any compact star with an exotic core,  is more challenging than for standard neutron stars because the properties of hadronic matter could rapidly change at the interface between the nuclear envelope and the  core.  In this case, the  Sturm-Liouville differential equation, see for example~\cite{1927ode..book.....I, 1971PhT....24j..52B}, describing  linear radial oscillations~\cite{1966ApJ...145..505B,1983tsp..book.....C}   should satisfy  appropriate boundary and interface conditions~\cite{1989A&amp;A...217..137H}. The main difficulty is to find a proper numerical procedure that  takes into account  that  the coefficients of the Sturm-Liouville differential equation can be discontinuous at the interface between nuclear and quark matter.  This problem has been discussed in General Relativity (GR)  in~\cite{VasquezFlores:2012vf, Pereira:2017rmp} for an
 hybrid stellar model with an envelop described by a Walecka model and the interior by quark matter. In this case  there exists a baryon and speed of sound  discontinuity at the interface between nuclear and quark matter.  Remarkably, the authors find that the last stable configuration does not coincide with the maximum stellar mass: the null frequency of  radial oscillations appears at central matter densities exceeding the central density of the maximum star. This difference is not due to  nonequilibrium processes, but it  instead depends on  the discontinuous behavior at the interface. Motivated by these results we  started to analyze the properties of radial oscillations, in particular  the effect of boundaries and  interface  discontinuities.  

In the present paper we propose a numerical algorithm based on an extension of the discretized Numerov's method that allows us  to properly describe the radial oscillations in GR, including the effect of   boundaries and discontinuous interfaces. We have  considered five different EoSs: three of them are based on microscopic physical models, while two EoSs are built to test the reliability of the numerical method in the presence of tunable interface discontinuities.  Our extended Numerov method  works with any considered background EoS and does not only provide  precise eigenfrequencies but also precise radial eigenfunctions.  We believe that our results can be of a certain interest because we  precisely deal with discontinuities and because we obtain hundreds of radial eigenfrequencies and eigenmodes  with a very high precision  using an algorithm that works on a laptop computer for just few tens of seconds. To show the reliability of the method we show the radial eigenmodes, finding that close to the boundaries  they have exactly the  behavior that can be inferred by  expanding the Sturm-Liouville equation. Moreover, we display the pressure oscillations, which in any considered case turn to be   continuous functions of the radial coordinate. Regarding the interface discontinuities,  we first consider  stellar configurations with a speed of sound discontinuity and then with both a matter density and speed of sound discontinuity and we determine the  spectrum of the radial oscillations.  In both  cases    we find that  the null mode appears at a  central density exceeding the one corresponding to the maximum mass. In other words, the last stable and the maximum stellar mass configurations do not coincide and therefore {\it twin} configurations may be realized.    Note that we assume that the radial fluctuations obey the same equation of state of the background.

The present paper is organized as follows. In Sec.~\ref{sec:background} we recall and discuss   the equation for the hydrostatic stellar equilibrium and we introduce the five EoSs that will be analyzed. In Sec.~\ref{sec:stationary} we revisit the equations of standing  radial oscillations, focusing on boundaries and interfaces. In Sec.~\ref{sec:numerical} we present our extended Numerov method for readily obtain the eigenfrequencies and the eigenmodes of the standing radial oscillations. The numerical results are shown in  Sec.~\ref{sec:Numerical_results}, where we perform as well a number of checks. We draw our conclusions in Sec.~\ref{sec:conclusions}.

\section{Background configuration}
\label{sec:background}
We assume a nonrotating spherically symmetric star with a Schwarzschild's line element
\be
ds^2 = e^{2\phi} dt^2 - e^{2\lambda}dr^2 - r^2(d\theta^2 +\sin^2\theta d\phi^2)\,,
\ee
where the metric potentials $\phi\equiv \phi(r)$ and  $\lambda \equiv \lambda(r)$ depend only on the radial coordinate.   
The  matter content is  treated as a perfect fluid with a barotropic EoS, $p\equiv p(\rho)$,  where $p$, and    $\rho$,  are respectively pressure and energy density.
For the radially symmetric time independent case,  the hydrostatic equilibrium is determined by the TOV's  differential equation
\begin{align}
\label{eq:TOV2}
p' &= (\rho+p)  \frac{m + 4 \pi p r^3}{2 m r -r^2}\,,
\end{align}
where the prime denotes the radial derivative  and 
\be\label{eq:mr}
m(r) = 4 \pi \int_0^r \rho(x) x^2 dx\,,
\ee is the gravitational mass within the spherical volume of radius $r$. For a given EoS and  central matter density, $\rho_c$, one can numerically integrate the TOV's equation and determine the stationary  configuration. The numerical integration begins at a small internal coordinate, $r_\text{min}$, and ends at the  stellar radius, $R$, corresponding to the  radial coordinate where the pressure  vanishes. The total gravitational mass is  then  $M = m(R)$. 

Once the TOV's equation have been solved, the metric functions are readily determined by
\begin{align}
e^{-2 \lambda} &= 1- 2 \frac{m}{r}\,,
\label{eq:lambda}
\end{align}
and by integrating
\begin{align}
 \phi' &= - \frac{p'}{\rho + p}  \,.
\label{eq:phi}
\end{align}
We also define the adiabatic speed of sound squared, the adiabatic index and  the adiabatic compressibility, respectively  by
\be
c_s^2 = \frac{\partial p}{\partial \rho}\,, \qquad \gamma = \frac{\rho + p}{p} c_s^2\,, \qquad 
\beta_s = \frac{1}{\rho c_s^2}\,, 
\label{eq:defs}
\ee
which determine the mechanical properties of matter. In particular, the adiabatic compressibility indicates  how stiff is the EoS, that is how difficult is to compress matter.  In any  microscopic EoS, $\beta_{s}(r)$  is a monotonically increasing function, however we shall also consider  stellar models with a non-monotonically decreasing adiabatic compressibility.

\subsection{Some general aspects of the TOV's configurations}

We remark some important aspects of the solutions of the TOV's equation. First of all, the TOV's equation determines the stationary stellar configurations, which could actually be unstable. To establish the stability  towards collapse or explosion one has to  study  the fluctuations on the top of the background stationary solution. 
The second point is that by expanding  the  TOV's equation close to the  stellar center one finds  that at the ${\cal O}(r^2)$ 
\begin{align}
p(r) &=p_c  - a_p r^2\,,\\ 
\rho(r) &=  \rho_c - a_\rho  r^2\,,\\ 
\phi(r) &= -|\phi_c| + a_\phi r^2\,,
\label{eq:taylor_center}
\end{align}
where the subscript $c$ indicates that the quantity evaluated at the stellar center and  $a_p$, $a_\rho$ and $ a_\phi$ are three positive quantities.
Finally, although $p(r)$ and $\phi(r)$ must be  continuous functions, the energy density and the speed of sound can be discontinuous. A discontinuity in $\rho$ is a possible consequence of a stellar onion structure where two subsequent layers  have different {\it chemical} composition, for example in the crust of standard neutron stars the matter density changes in a slightly discontinuous way due to the fact that different nuclei are energetically favored.  The discontinuity in $c_s^2$ is instead characteristic of an interface between materials with different mechanical properties, for example  the speed of sound can abruptly change because of the presence of a crystalline phase. A large matter density and speed of sound change  can be realized in hybrid stars  at the interface between between nuclear  and deconfined quark matter or within quark matter at the interface between the color-flavor locked phase and the crystalline color superconducting phase, see~\cite{Anglani:2013gfu} for a review.

From Eqs.~(\ref{eq:TOV2},\ref{eq:mr},\ref{eq:lambda},\ref{eq:phi}) and (\ref{eq:defs}) it follows that \be
\text{discontinuos  }  \rho \Rightarrow
\begin{cases}
p' & \text{discontinuous}\\
m' & \text{discontinuous}\\
\phi' & \text{continuous}\\
\lambda' & \text{discontinuous}\\
\rho' & \text{delta function}\\
\end{cases}
\label{eq:rhodisc}
\ee
while a  discontinuous speed of sound implies that only the derivative of the energy density is discontinuous; more precisely 
\be
\text{discontinuos  }  c_s \Rightarrow
\begin{cases}
p' & \text{continuous}\\
m' & \text{continuous}\\
\phi' & \text{continuous}\\
\lambda' & \text{continuous}\\
\rho' & \text{discontinuous}
\label{eq:csdisc}
\end{cases}
\ee
therefore, loosely speaking, an EoS with a discontinuous speed of sound has a mild discontinuity. Clearly, these are not real discontinuities: they should be understood as rapid radial variations,  on a length scale much smaller that $R$.   

\subsection{The equations of state}
\label{sec:EoS}

\begin{figure}[h!]
\includegraphics[width=0.45\textwidth]{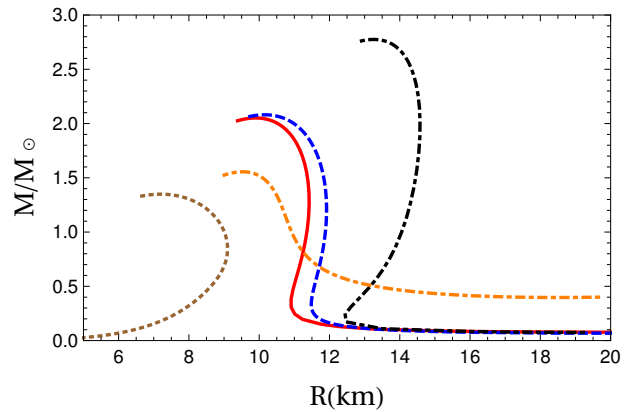}
\caption{Mass-radius diagram obtained with  five different EoSs. The solid red line corresponds to the SLy4~\cite{Douchin:2001sv}, the dashed blue line to the BL~\cite{Bombaci:2018ksa}  and the black dashed dotted line to the MS1~\cite{Mueller:1996pm} EoSs.  The brown dotted line corresponds to the piecewise polytropic defined in Eq.~\eqref{eq:piecepoly}, see also the discussion below;  the orange dashed dotted line corresponds to the hybrid star model defined in Eq.~\eqref{eq:hybrid}. }
\label{fig:MR}
\end{figure}

The EoS is a necessary ingredient for the determination of the equilibrium stellar configuration and should be determined microscopically, taking into account the relevant degrees of freedom and interactions. However, nuclear interactions above the nuclear saturation density are not well known and therefore the EoSs at large densities are obtained by extrapolation. For this reason, there is a large number of possible  EoSs. We shall restrict to  five different cases: We will consider three EoSs that have been derived by some plausible microscopic modeling: 
the SLy4~\cite{Douchin:2001sv}, the BL~\cite{Bombaci:2018ksa} and the MS1~\cite{Mueller:1996pm} EoSs,  a piecewise polytropic and an hybrid EoS. 
Upon inserting each of these EoSs in Eq.~\eqref{eq:TOV2} one obtains the   mass radius diagram  reported in Fig.~\ref{fig:MR}, which represents the gravitational mass versus radius  obtained  by changing the central density. The solid red line corresponds to the results obtained with the SLy4, the dashed blue line to the BL and the black dashed dotted line to the MS1. All of  these three EoSs have a maximum mass exceeding the observational bound \cite{Demorest:2010bx,Antoniadis:2013pzd}, see also~\cite{Rezzolla:2017aly}, of $2 M_\odot$, where  $M_\odot$ is the solar mass.

Next, we consider two EoSs built to study  the effect of thermodynamic discontinuities.
The  piecewise polytropic EoS is defined as
\be\label{eq:piecepoly}
p = \begin{cases} K_1 \rho^{\Gamma_1} & \text{for  } \rho < \rho_t\\
K_2 \rho^{\Gamma_2}& \text{for  } \rho > \rho_t\\
\end{cases}\,,
\ee
and we assume a transition density $\rho_t\simeq 2.98 \rho_\text{sat}$, where the saturation density and pressure are respectively  
$\rho_\text{sat}  \simeq 2.7 \times 10^{14} \text{ g cm}^{-3}$  and $p_\text{sat} \simeq 4 \times 10^{33} \text{ dyne cm}^{-2}$. In principle, $K_1$ and $K_2$ as well as $\Gamma_1$ and
$\Gamma_2$ are the parameters describing the properties of  matter in two different phases, see for example~\cite{Shapiro:1983du}. Here we employ a simplified approach aimed to reproduce  the nuclear saturation point and based on the assumption that the matter density is continuous. In this way   we obtain that 
\be\label{eq:K1_K2}
K_1=  \frac{p_\text{sat}}{\rho_\text{sat}^{\Gamma_1}}\,, \qquad K_2 = \frac{p_\text{sat}}{\rho_\text{sat}^{\Gamma_2}}\,,
\ee
 At the interface, the   speed of sound and the compressibility discontinuities  can be expressed as
\begin{align}
\Delta c_s^2 & =(c_{s2}^2-c_{s1}^2) \vert_t=  (\Gamma_2-\Gamma_1) \frac{p_t}{\rho_t}\,,\\
\Delta \beta_s &= (\beta_{s2}^2-\beta_{s1}^2) \vert_t = \frac{\Gamma_1-\Gamma_2}{\Gamma_1\Gamma_2} \frac1{p_t}\,,
\end{align}
where $c_{s1}$ and $c_{s2}$ are the speeds of sound in the two phases.
In this case one can even probe configurations with a nonmonotonic compressibility. In particular, we considered the model characterized by  $\Gamma_1=2$, $\Gamma_2=3$, as well as $\Gamma_1=3$, $\Gamma_2=2$ that has   a speed of sound jump $\Delta c_s = +0.05$ at the transition point . The   TOV solutions for the latter case correspond to the  dotted brown line in Fig.~\ref{fig:MR}. In this case the maximum mass is  $M\simeq1.24 M_\odot$ and the corresponding radius $R \simeq 6.6$ Km, for  a central density $\rho_c\simeq 25.4 \rho_\text{sat}$. The small values of mass and radius and the large value of the critical density are due to the fact we have considered a model in which part of the inner part is softer than the outer part.


Finally, we consider the hybrid star model with  quark matter in the interior and an envelope described by a polytrope:
\be\label{eq:hybrid}
p = \begin{cases}
K \rho^{\Gamma}& \text{for  } p < p_t\\
 c_{sq}^2 (\rho - 4 B ) & \text{for  } p > p_t\\
\end{cases}
\ee
where $K=p_\text{sat}/\rho_\text{sat}^\Gamma$ (we tried to  different values of $K$, with qualitatively the same results), $c_{sq} $ is the speed of sound of quark matter and we take the bag constant $B=(165 \text{MeV})^4$.   For the envelope we  take $\Gamma=4/3$, as appropriate for a nonrelativistic electron gas.  Unless differently stated, we use $c_{sq}^2=1/3 $.  At the transition point 
\be
K \rho_1^{\Gamma} = c_{sq}^2 (\rho_2 - 4 B )\,, 
\ee
and the energy density discontinuity can be expressed as
\be\label{eq:deltarhot}
\Delta\rho_t=\rho_2 -\rho_1 = \frac{K \rho_1^{\Gamma}}{c_{sq}^2} + 4 B -\rho_1\,,
\ee
and therefore
\be
\frac{\partial\Delta\rho_t}{\partial \rho_1} = \frac{c_{se}^2-c_{sq}^2}{c_{sq}^2} = \frac{\Delta c_{s}^2}{c_{sq}^2}\,,
\ee
where $c_{se}^2= \Gamma p/\rho$ is the speed of sound squared in the envelope. In this case both $\Delta \rho_t$ and $\Delta c_s$ are nonzero.
The  corresponding TOV solution is reported in Fig.~\ref{fig:MR} by a orange dashed dotted line for 
 $\rho_2 = \rho_1$.

In principle one may consider more refined hybrid  star models, see for example~\cite{Alford:2004pf, Chen:2015mda},  or piecewise polytropes, see~\cite{Shapiro:1983du}, however for our purposes it is enough to consider  these relatively simple stellar  models. Indeed, these models have the basic ingredients to test the properties of radial fluctuations in the presence of  tunable speed  of sound and/or  energy density discontinuities. We  notice that both models  have a maximum mass that  is below  the observed $2 M_\odot $ bound~\cite{Demorest:2010bx,Antoniadis:2013pzd, Rezzolla:2017aly}, however this is irrelevant for our purposes, because these models  simply serve  to test the numerical Numerov's method and to study the effect of tunable discontinuities.

\section{Standing radial oscillations}
\label{sec:stationary}
Once the TOV stationary configuration has been determined, one can probe its stability towards collapse or explosion by considering  a small harmonic radial perturbation of the form
\be\label{eq:stationary}
\delta r = X(r) e^{i \omega t}\,,
\ee
where  $X(r)$ and  $\omega$ are respectively  the amplitude and the frequency of the standing wave. The TOV stationary configuration is unstable if some stellar mode has an imaginary frequency. The conditions for the existence of standing waves are discussed in~\cite{1983tsp..book.....C}; here we shall restrict to adiabatic  oscillations that conserve  the total baryonic number that are  slow with respect to the microscopic dynamics, see the discussion in~\cite{1989A&amp;A...217..137H}. 
Within these restrictions,  the linearized perturbation  equations     can be written as a second order homogenous differential equation
\begin{align}\label{eq:X}
 \omega^2 e^{2 (\lambda-\phi)}X =-c_s^2 X'' - \left((c_s^2)'-Z + 4\pi r \gamma p e^{2\lambda} -\phi'\right)X' \nonumber \\ - \left(2 (\phi')^2 + \frac{2 m}{r^3}e^{2\lambda} - Z' + 4\pi (p+w) Z r e^{2 \lambda}  \right)X\,,
\end{align}
where 
\be
Z= c_s^2\left(\phi' - \frac{2}r\right)\,,
\ee 
while  $\gamma$ and $c_s$ have been defined in Eq.~\eqref{eq:defs}, and therefore have the same value determined for the background. Following~\cite{1966ApJ...145..505B}, we redefine the radial displacement as
\be\label{eq:xidef}
\xi = X r^2 e^{-\phi}\,,
\ee
in this way the differential Eq.~\eqref{eq:X}  can be rewritten as a Sturm-Liouville differential equation
\be
(H \xi')'=-(\omega^2 W + Q)\xi\,,
\label{eq:sturm-liouville}
\ee
where
\begin{align}
H &=  r^{-2} (\rho + p)  e^{\lambda + 3 \phi} c_s^2 \nonumber \\
Q &=  r^{-2} (\rho + p) e^{\lambda + 3 \phi} (\phi'^2 + 4 r^{-1}\phi'-8 \pi e^{2 \lambda}p) \nonumber  \\
W &= r^{-2} (\rho + p)  e^{3\lambda +  \phi}\,,
\label{eq:coefficinets_SL}
\end{align}
are the relevant background functions. This is a particularly convenient expression because the  
 Lagrangian fluctuation of the pressure takes the simple expression 
\begin{align}
\Delta P &= -  c_s^2 (p+\rho) r^{-2} e^\phi \, \xi'\,, \label{eq:DeltaP}
\end{align}
and because the solution of Eq.~\eqref{eq:sturm-liouville} are known to be discrete: we will indicate with  $\hat\xi_n$  the eigenmodes with   $n$ nodes and with $\omega_n$ the corresponding  frequency. 
The actual determination of the  solution requires the study of the full numerical problem, including the appropriate boundary conditions.


%

\subsection{Boundary conditions and interfaces} 
The differential Eq.~\eqref{eq:sturm-liouville} with the boundary conditions
\begin{align}\label{eq:sturmian_1}
\alpha_1 \xi(0) + \alpha_2 \xi'(0) &=0 \qquad \alpha_1^2+\alpha_2^2>0\,, \\
\label{eq:sturmian_2}
\beta_1 \xi(R) + \beta_2 \xi'(R) &=0 \qquad \beta_1^2+\beta_2^2>0\,,
\end{align}
forms a Sturmian system and it can be proved, see for instance~\cite{1927ode..book.....I}, that $\omega_n^2$ are real and ordered \be
\omega_n^2 < \omega_{n+1}^2\,,
\ee
meaning that the fundamental $0$th mode has the lowest frequency. For radial oscillations the boundary condition at the stellar center is 
\be
\xi(0) =0\,, 
\ee
because the stellar center cannot be displaced, therefore it is equivalent to the condition in Eq.~\eqref{eq:sturmian_1} with $\alpha_2=0$. 
However, the boundary condition at the stellar surface is not in general as in Eq.~\eqref{eq:sturmian_2}. The reason is that  expanding the Sturm-Liouville equation close to the stellar surface one obtains, see for example~\cite{1966ApJ...145..505B}, 
\be\label{eq:expansionR}
\xi'_n = \gamma^{-1} R^{-1}(4 + e^{2 \lambda} M/R  + \omega_n^2 e^{-2 \phi} (R^3/M))\xi_n\,,
\ee
which is similar to Eq.~\eqref{eq:sturmian_2}, but with the important difference that the  Sturmian boundary condition does not  depend on $\omega_n$.  The only case in which Eq.~\eqref{eq:expansionR} turns in Eq.~\eqref{eq:sturmian_2} with $\beta_2=0$ is when $\gamma$ diverges at the stellar surface, as for strange stars. In this case $r=R$ is a regular point and all the solutions of  Eq.~\eqref{eq:sturm-liouville} are regular at the stellar surface. If $\gamma$ is finite then $r=R$ is a regular singular point and there is an unphysical  diverging solution. 

These considerations led us to quest what are the general boundary conditions to be used for the regular solution of Eq.~\eqref{eq:sturm-liouville}. Certainly, since it is an  homogeneous differential equation  the absolute value of the eigenfunction is irrelevant. As usual, we will use this freedom to set $\xi(R) =1$, in arbitrary units.  To eliminate the diverging  solution we show that it is enough to specify the boundary condition close to  the stellar center where  Eq.~\eqref{eq:sturm-liouville} admits two possible solutions 
\be\label{eq:xiclose}
\xi = C_1 ( r^3 + {\cal O}(r^{5})) + C_2 (1+ b r^2 + {\cal O}(r^{4}))  \,,
\ee
where  $C_1$ and $C_2$ are integration constants and $b$ depends on the background. 
Since the fluid at the stellar center cannot be displaced by a radial oscillation it follows that one has to take  $C_2=0 $  to eliminate  the unphysical solution. In our numerical solution we will require that 
 \be\label{eq:correct}
\hat \xi_n \propto r^3\,,
\ee
 and  technically this  will be done  imposing the boundary conditions 
\be\label{eq:bound0}
\xi(r_\text{min})= C_1 r_\text{min}^3 \quad \text{and} \quad \xi'(r_\text{min})=3 C_1 r_\text{min}^2\,,
\ee
where $r_\text{min}$ is the smallest radial distance considered in the integration of the TOV's equation. Therefore, we take  both boundary conditions  close to  the origin and we do not need an extra boundary condition at the surface.
Let us insist on this aspect:  the only boundary conditions that we impose to solve the differential equation are close to  the stellar center. It is sometimes stated that one should impose a  boundary condition at the stellar surface to avoid the unphysical diverging solution  and thus enforce $\Delta P(R)=0$. 
Instead in our approach, the vanishing of the pressure fluctuations   at the stellar surface is a consequence of the boundary conditions at the origin.   In other words, any physical solution of the Sturm-Liouville differential equation with the correct boundary condition at $r=r_\text{min}$ does  automatically satisfy the requirement that $\Delta P (R)=0$. 

Regarding the radial dependence of $\Delta P$ it is maybe interesting to add few remarks.  Since close to the stellar center the displacement field behaves as in Eq.~\eqref{eq:xiclose}, the pressure oscillation at the stellar center is well defined and  given by
\be\label{eq:DeltaPc}
\Delta P_c = -3 a_\xi c_{s,{c}}^2 (p_c+\rho_c)e^{\phi_{c}}\,,
\ee  
where, as in Eq.~\eqref{eq:taylor_center}, we have indicated with the subscript $c$ the values of the functions at the stellar center. Clearly $\Delta P_c$ is extremely small,   because $\phi_{c}$ is large and negative. By increasing $r$ we know that  the various quantities change as in Eq.~\eqref{eq:taylor_center}, in particular
$\phi(r)$ increases  and the pressure oscillation  exponentially grows. We also know that the  pressure fluctuation vanishes at the stellar surface this means that $\Delta P(r)$ has  at least a  maximum, or a minimum. In Sec.~\ref{sec:numerical}  we shall see that our numerical procedure is in agreement with this outlined behavior and the extremum of $\Delta P(r)$ is located very close to the stellar center.  

We now turn to the interface between different stellar layers. The  continuity of  $\Delta P$ ensures that the system is always close to equilibrium.  Since  $\Delta P$ is a function of $\rho$ and $c_s^2$ it seems that any discontinuity of these quantities could produce a  pressure jump. But this is unphysical, unless the perturbation arises in a time scale much shorter than the typical  equilibrium timescale, which is not the case for  the slow  oscillations considered in the present article.  Since $\phi$ is always a continuous function,  from Eq.~\eqref{eq:DeltaP} we have that the continuity of the pressure perturbation implies  that  if  $c_s^2 (\rho +p)$ is discontinuous at a  radial coordinate  $\bar r$,   then $\xi'$ is  discontinuous in $\bar r$.  
More precisely, we can say that if there is a shell of negligible depth, $\delta$, centered at $\bar r$   where $c_s^2 (\rho +p)$  abruptly changes,  then  labeling with the  $I$ ($E$)  the quantities evaluated at the internal side (respectively external) side of the boundary, the continuity of the displacement and of the pressure imply that
\begin{align}
 \xi_I= \xi(\bar r- \delta) & \simeq \xi(\bar r+ \delta) = \xi_E\,, \nonumber\\
 c_s^2  (\rho +p) \xi'\vert_I & \simeq  c_s^2 (\rho +p) \xi'\vert_E \,, 
\label{eq:left_right}
\end{align} 
meaning that  $\xi$ is a continuous function and   although $\xi'$ is discontinuous in $\bar r$, the combination $c_s ^2(p +\rho) \xi'$ is always a  continuous function.


\section{The numerical method}
\label{sec:numerical}
We have developed an extended  Numerov discretization method for the solution of the Sturm-Liouville equation \eqref{eq:sturm-liouville} that  takes into account the  appropriate boundary conditions in Eq.~\eqref{eq:bound0}, as well as the possible speed of sound and density discontinuities in Eq.~\eqref{eq:left_right}. The method  consists in discretizing   the radial coordinate in $N$ steps transforming the Sturm-Liouville differential equation in an eigenvalue problem~\cite{1983MNRAS.202..159G}. The advantage of our   procedure with respect to other methods, see for example~\cite{1966ApJ...145..505B,Kokkotas:2000up},   is that it simultaneously provides  many radial frequencies and eigenmodes, and no unphysical solution appears if one properly imposes the boundary conditions in Eq.~\eqref{eq:bound0}. Moreover, there is a number of  check that can be used to test the convergence of the method. 

The proposed extended Numerov method works as follows. The  Sturm-Liouville differential equation, see Eq.~\eqref{eq:sturm-liouville},   can be written as
\be\label{eq:differential}
A_1 \xi^{''} +A_2 \xi^{'}+ A_3  \xi = \omega^2 \xi\,,
\ee
where $A_1, A_2$ and $A_3$ can be obtained by expanding Eq.~\eqref{eq:sturm-liouville} using the coefficients in Eq.~\eqref{eq:coefficinets_SL}. Then we discretize the radial coordinate as
\be\label{eq:rn}
r_n = n \epsilon \,,  
\ee
where $\epsilon= R/N$ and $n=1,\dots, N$. We define $ \xi({r_i}) = \xi_i$ and $A_{1n}=A_{1}(r_{n})$, $A_{2n}=A_{2}(r_{n})$ and $A_{3n}=A_{3}(r_{n})$ 
in such a way that Eq.~\eqref{eq:differential} turns in
\be\label{eq:differential_discretized}
A_{1n} \xi^{''}_n +A_{2n} \xi^{'}_n+ A_{3n}  \xi_n = \omega^2 \xi_n\,,
\ee
and then we can cast Eq.~\eqref{eq:differential_discretized} as the eigenvalue problem
\be\label{eq:eigeneq}
  A \bm \xi = \omega^2 \bm \xi\,,
\ee
where  $ \bm \xi^t = (\xi_1, ..., \xi_{N})$ and $A$ is the  matrix obtained 
by expressing the radial derivatives as finite differences.  We can discretize the radial derivatives  at any desired order;  we checked that the method works  considering the lowest order expansion of the derivatives but we used
\begin{align}
\xi'_n &=  \frac{\xi_{n-2}-8 \xi_{n-1}+8 \xi_{n+1}-\xi_{n+2}}{12 \epsilon} + {\cal O} (\epsilon^4)\nonumber \\
\xi^{\prime \prime}_n &= \frac{-\xi_{n-2}+16 \xi_{n-1}-30 \xi_n+16 \xi_{n+1}-\xi_{n+2}}{12 \epsilon^2}+ {\cal O} (\epsilon^3)\,,
\label{eq:discrete_der}
\end{align}
where $\xi'_n \equiv \xi'(r_n)$ and $\xi''_n \equiv \xi''(r_n)$.
The discretized version of Eq.~\eqref{eq:differential} for $2<n<N-2$ is
\begin{align}
\label{eq:discretized}
\omega^2 \xi_n &= {\xi_{n-2}} \left(\frac{{A_{2n}}}{12 \epsilon}-\frac{{A_{1n}}}{12 \epsilon^2}\right)+{\xi_{n-1}} \left(\frac{4 {A_{1n}}}{3 \epsilon^2}-\frac{2 {A_{2n}}}{3 \epsilon}\right)\nonumber	\\ &+ \xi_n \left({A_{3n}}-\frac{5 {A_{1n}}}{2 \epsilon^2}\right) + \xi_{n+1} \left(\frac{4 {A_{1n}}}{3 \epsilon^2}+\frac{2 {A_{2n}}}{3 \epsilon}\right)\nonumber \\ &+{\xi_{n+2}} \left(-\frac{{A_{1n}}}{12 \epsilon^2}-\frac{{A_{2n}}}{12 \epsilon}\right) = \sum_{m=n-2}^{n+2} a_{nm} \xi_m \,,  
\end{align}
which defines the matrix entries $a_{nm}$ for $2<n<N-2$ and $2<m<N-2$. All the other matrix elements are, for the time being, zero, indeed one cannot use the above definitions for the first two and last two rows of $A$.

\subsection{Boundaries}
\label{sec:boundaries}
We first discuss how to implement the boundary condition close to the stellar center,  properly defining the first two rows of the matrix $A$ to impose Eq.~\eqref{eq:bound0}.  
We recall that the implementation of  these condition is extremely important to avoid the unphysical solution, which is proportional to $C_2$ in Eq.~\eqref{eq:xiclose}, and to quantize the    eigenfrequencies. 
In agreement with  Eq.~\eqref{eq:bound0}, close to the stellar center the first two discretized values of any eigenmode should be 
\be\label{eq:bound01}
\xi_1 = C_1 r_\text{min}^3 \qquad \text{and} \qquad \xi_2 = C_1 (r_\text{min} +\epsilon)^3\,,
\ee
where $r_\text{min}$ is the minimum value considered in the numerical integration of the TOV's equation.
Then, we require that 
\begin{align}
\omega^2   \xi_1 &= a_{12}  \xi_2\,,\nonumber\\
\omega^2  \xi_2 &= a_{21}  \xi_1\,,
\label{eq:a12}
\end{align}
which is the simpler way to link the first two values of the displacement close to the stellar center. Obviously,  we do not know  $\omega$, therefore it seems that we cannot fix the   values of $a_{12}$ and $a_{21}$. However, from the above equations we obtain that  $a_{12} a_{21} =\omega^4$, and
\be
\frac{a_{12}}{a_{21}} =\left( \frac{r_\text{min}}{r_\text{min} +\epsilon}\right)^6\,,
\ee
which determines the ratio between these two matrix elements. Suppose that   we fix the eigenmode,  for simplicity  $\omega^{4}= 1$, then   we have that
 \be a_{12}= \frac{1}{a_{21}} = \left( \frac{r_\text{min}}{r_\text{min} +\epsilon}\right)^3\,,
\ee 
and thus these matrix elements are now fixed.
Now, if we  define the  top left corner of  $A$ using Eq.~\eqref{eq:a12}, therefore  as the block  matrix
\be
\left( \begin{array}{cc}
0 & a_{12} \\
a_{21} & 0 \end{array} \right)\,,
\label{eq:topleftA}
\ee
we are  actually imposing that any eigenfunctions has the boundary conditions of Eq.~\eqref{eq:bound01}. This will result in  two spurious   eigenvalues  $\omega^{2}=\pm 1$ in the  spectrum. In the end, since we know the values of  these two spurious eigenvalues, we can easily identify and remove them as well as the corresponding eigenvectors.  

Regarding the stellar surface, we do not impose any boundary condition.  To define the matrix elements at the stellar surface, or more precisely   the rows $N-1$ and $N$ of the $A$ matrix, we cannot use the discretization of the first and second derivatives of Eq.~\eqref{eq:discrete_der}. The reason is that  the background  quantities are not defined for   $r>R$ and therefore the $\xi_{N+1}$ and $\xi_{N+2}$ elements are unphysical. We tried different extrapolation method, which however lead us to  different  results. Thus we  redefine the derivatives close to the $r=R$ boundary as  
\begin{align}
\xi'_{N-1} &=  \frac{-\xi_{N-4}+6 \xi_{N-3}-18 \xi_{N-2}+3 \xi_{N-1}+10 \xi_{N} }{12 \epsilon} \nonumber \\
\xi'_{N}&=  \frac{\xi_{N-4}-16/3 \xi_{N-3}+12 \xi_{N-2}-16 \xi_{N-1}+25/3 \xi_{N} }{4 \epsilon} \nonumber \\
\xi''_{N-1} &=  \frac{-\xi_{N-4}+4 \xi_{N-3}+6 \xi_{N-2}+11 \xi_{N-1}-20 \xi_{N} }{12 \epsilon} \nonumber \\
\xi''_{N} &=  \frac{11 \xi_{N-4}-56 \xi_{N-3}+114 \xi_{N-2}-104 \xi_{N-1}+35 \xi_{N} }{12 \epsilon} 
\label{eq:rightderivatives}
\end{align}
meaning that for $n=N-1$ and $n=N$ we have that
\begin{align}
\omega^2 \xi_n = \sum_{m=N-4}^{N}a_{nm}\xi_m\,,
\end{align}
where the $a_{nm}$ coefficients for $ N-1\leq n \leq N$ and $N-4 \leq m \leq N$ can be determined inserting the   Eqs.~\eqref{eq:rightderivatives} in  Eq.~\eqref{eq:differential_discretized}.

\subsection{Interfaces}
\label{sec:sound_discontinuity}

We now consider how to discretize the differential equation close to  $\bar r$, corresponding to  the interface where  the speed of sound  and/or the matter density are discontinuous. The continuity of the  pressure oscillation in Eq.~\eqref{eq:DeltaP} implies that $\xi'(\bar r)$ is discontinuous and that   $\xi''(\bar r)$ is a Dirac delta function. One therefore needs to isolate the discontinuity and properly expand on the left and on the right of $\bar r$. We remark that in any case it is important to obtain an expression that is ``symmetric" around the discontinuity. In principle one is tempted to define the left-derivative for $r<\bar r$ and the right derivative for $r>\bar r$, but this method does not work with the Numerov discretization. The reason is that in this way one would obtain a block diagonal matrix, with a separation between interior  modes  and exterior modes. One could connect the left and right derivatives by inserting an additional intermediate point, however we found a better and faster way to deal with the discontinuity.

First, we isolate the discontinuous point: For any $N$  we build the set $r_1, \dots, r_N$ according to Eq.~\eqref{eq:rn} and we select $k$ such that $\bar r-r_k >0$ is a minimum and  we  define  $\delta < \epsilon$ asking that     $r_k +\delta > \bar r$ and   $r_{k+1} -\delta < \bar r$. Then, we use the fact that $c_s ^2(p +\rho) \xi'$ is a continuous function of $r$, see Eq.~\eqref{eq:left_right}.  
It follows  that for any $r$ in the neighborhood  of $\bar r$
defined as  $|r - \bar r| < \delta$ 
\begin{equation}
\xi^\prime S \vert_{r-\delta}\simeq\xi^\prime S \vert_r\simeq \xi^\prime S \vert_{r+\delta}\,,
\label{eq:continuity}
\end{equation} 
where the scaling function is defined as
\be
S(r) \equiv c_s ^2(p +\rho)\,,
\ee
and  the value of the sound speed and of the energy density in $r$ depends on whether $r$ is smaller or bigger than $\bar r$. By expanding the   displacement function on the left and on the right of  $\bar r$, multiplying  by the scaling function and taking into account Eq.~\eqref{eq:continuity},  we  can express the first and the second  derivative  in the symmetric forms
\begin{align}
\xi'_k &= \frac{S(r_k +\delta) (\xi_{k+1}-\xi_k)+S(r_k -\delta) (\xi_k-\xi_{k-1})}{2 \epsilon S(r)}\nonumber\\
\xi''_k &= \frac{S(r_k +\delta) (\xi_{k+1}-\xi_k)-S(r_k -\delta) (\xi_k-\xi_{k-1})}{\epsilon^2 S(r)}\,,
\label{eq:derivatives_disc2}
\end{align}
where $\xi_k = \xi(r_k)$, $\xi'_k = \xi'(r_k)$ and  $\xi''_k = \xi''(r_k)$ and we have used the leading order expansion of the symmetric derivative. 
As a check, if $S(r)$ is continuous in $r$, the above expressions give the standard discretized definition of the first and second derivatives.  Upon substituting Eq.~\eqref{eq:derivatives_disc2} in Eq.~\eqref{eq:differential} one obtains the matrix elements $ a_{lm}$ with $l=k, k+1$ and $m=l-1,l,l+1$.  Therefore, around the discontinuous interface the matrix equation \eqref{eq:eigeneq} can be written as 
\be
\omega^2 \xi_l =  \sum_{m=l-1}^{l+1} a_{lm} \xi_m \,,
\ee
and thus, taking into account all the above discussion, the  form of the $A$ matrix in Eq.~\eqref{eq:eigeneq} turns to
\begin{widetext}
\begin{equation}A=\left(\begin{array}{cccccccccccccc}
0 & a_{12} & 0 & 0 & 0 & 0 &0  &  \cdots  & 0 & 0 & 0 & 0 & 0 & 0\\ 
 a_{21} & 0 & 0 & 0 & 0 & 0 & 0 &  \cdots & 0 & 0 & 0 & 0 & 0& 0 \\ 
 a_{31} & a_{32}  & a_{33} & a_{34} & a_{35}& 0  & 0 &  \cdots  & 0 & 0 & 0 & 0 & 0& 0 \\ 
 0 & a_{42} & a_{43} & a_{44} & a_{45} & a_{46} & 0 & \cdots& 0  & 0 & 0 & 0 & 0 & 0 \\ 
\vdots & &&  &  & & \ddots &  & &   & & & & \vdots  \\ 
\vdots & &&  & &  &a_{k k-1}  & a_{k k} & a_{k k+1} &   & & & & \vdots  \\ 
\vdots & &&  & &  &  & a_{k+1k} & a_{k+1 k+1} & a_{k+1 k+2}  & & & & \vdots  \\ 
\vdots & &&  &  & & &  & &\ddots   & & & & \vdots  \\ 
0 & 0 & 0 & 0 & 0  & 0 & 0 &  \cdots  &   a_{N-3N-5}& a_{N-3 N-4} & a_{N-3 N-3}  & a_{N-3 N-2} & a_{N-3 N-1}  & 0 \\ 
0 & 0 & 0 & 0 & 0  & 0 & 0 &  \cdots  & 0 &   a_{N-2 N-4}& a_{N-2 N-3} & a_{N-2 N-2}  & a_{N-2 N-1} & a_{N-2 N} \\ 
0 & 0 & 0 & 0 & 0  & 0 & 0 & \cdots & 0 & a_{N-1N-4} &  a_{N-1 N-3}& a_{N-1 N-2} & a_{N-1 N-1} & a_{N-1 N} \\ 
0 & 0 & 0 & 0 & 0  & 0 & 0 & \cdots & 0 &  a_{N N-4} &  a_{N N-3}& a_{N N-2} & a_{N N-1} & a_{N N}
\end{array}\right)\nonumber \end{equation}
\end{widetext}
where the first two lines and the last two lines have the peculiar form  determined in Sec.~\ref{sec:boundaries} to describe the stellar center and surface, respectively. The  two central lines have to be inserted to take into account the interface discontinuities. In principle, one can consider an EoS with an arbitrary  number of discontinuities: the corresponding $A$ matrix would then be a generalization of the one shown above, with an additional matrix element   for each interface discontinuity.

\section{Numerical results and checks}
\label{sec:Numerical_results}

Here we report the results obtained with the extended Numerov method using  the method presented in the previous section and the  EoSs discussed in Sec.~\ref{sec:background}. We will first analyze the microscopic EoSs, characterizing the radial displacement and pressure and doing a number of numerical checks. Then we turn to the EoSs with tunable parameters. 

\subsection{Microscopic equations of state}

\begin{figure}[h!]
\includegraphics[width=0.45\textwidth]{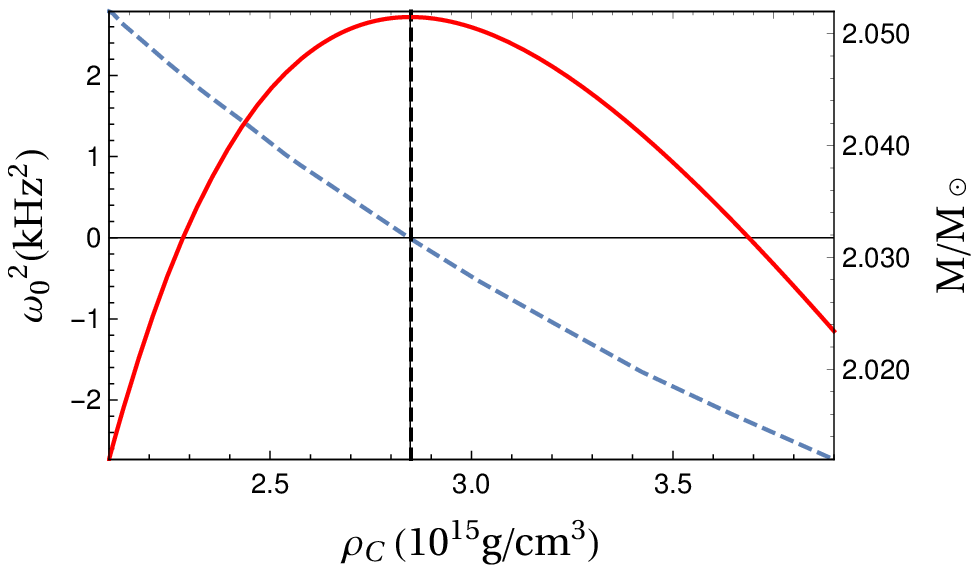}\qquad
\includegraphics[width=0.45\textwidth]{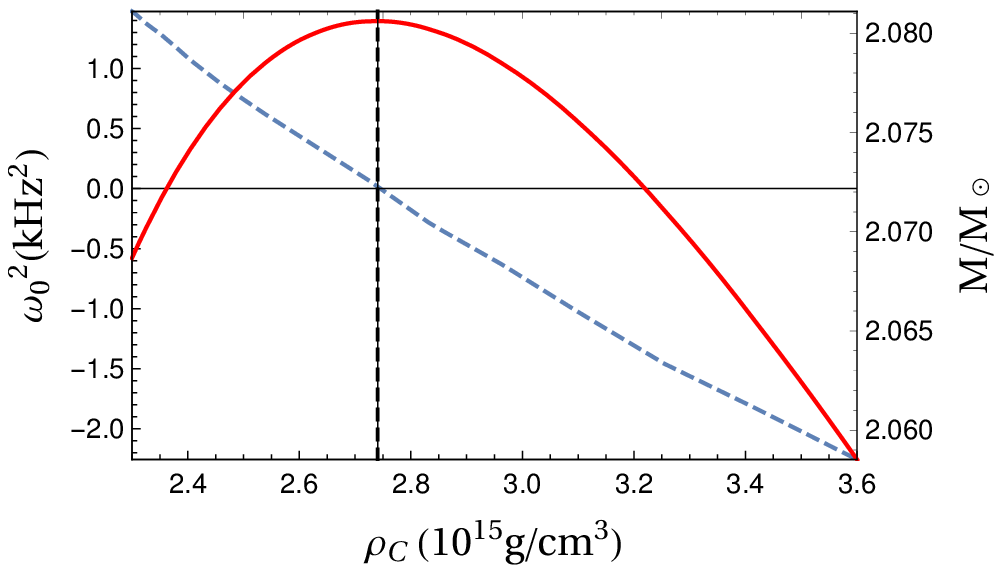} 
\includegraphics[width=0.45\textwidth]{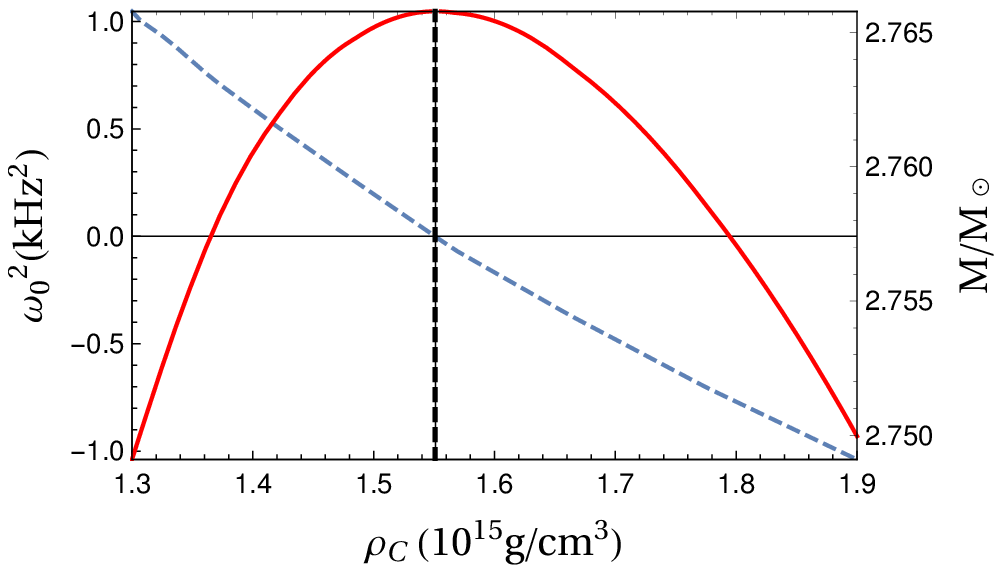}
\caption{Masses and fundamental eigenfrequencies as a function of the stellar central density for three different EoSs, from  top to  bottom: SLy4, BL, MS1, see Sec.~\ref{sec:background} for more details.
 The solid red lines correspond to the mass, in units of the solar mass, the dashed blue lines correspond to the  frequency squared of the fundamental mode. The null mode appears when the dashed blue line intersects the horizontal axis, which is exactly the same central density where the gravitational mass reaches the maximum value. The vertical  dashed line indicates  the  null mode while the solid vertical line indicates the maximum stellar mass.  These two lines perfectly  overlap. }
\label{fig:Omega_SLY}
\end{figure}
In Fig.~\ref{fig:Omega_SLY} we show the mass and the fundamental eigenfrequency as a function of the central density for the three microscopic EoSs discussed in  Sec.~\ref{sec:background}.  For these three cases the last stable configuration, corresponding to the null mode,  coincides with the maximum mass. 

\begin{figure}[h!]
\includegraphics[width=0.45\textwidth]{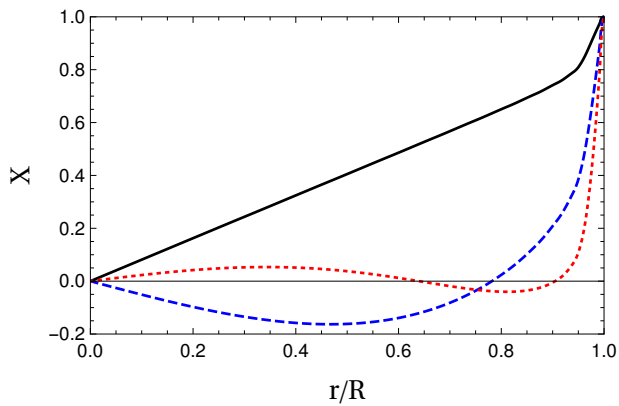}
\caption{First three radial eigenmodes as a function of the radial coordinate obtained using  the SLy4 EoS at  $\rho_c=0.98 \times 10^{15}$ g/cm$^3$ by the extended Numerov method with $N=500$ discretized points. The displacement has been normalized to $1$, in arbitrary units, at the stellar surface. In agreement with Eq.~\eqref{eq:XNrs},  close to the stellar center the displacements are linearly dependent on the radial coordinate. In agreement with Eq.~\eqref{eq:XpR}, close to the stellar surface the derivative of the displacement increases with $n$. }
\label{fig:radial_1}
\end{figure}

Regarding the radial eigenfunctions, we find that those obtained with the microscopic EoSs are similar, thus we only show  the results obtained with the SLy4 EoS.  In Fig.~\ref{fig:radial_1} we report the first three radial eigenmodes obtained with the SLy4 EoS for $\rho_c=0.98 \times 10^{15}$ g/cm$^3$, corresponding to a star with  mass $M\simeq 1.4 M_\odot $ and radius $R \simeq 11.4$ km,    obtained  with $N=1500$ discretized points. The obtained curves are smooth and we checked that the interpolated functions and the corresponding eigenfrequencies are  solutions of the differential equation governing the radial fluctuations with an error of the order of few percent. An important   nontrivial  check is that the numerically obtained radial displacements have the correct behavior at the boundaries, that is close to the stellar center and  the stellar surface.  Sufficiently close to the stellar center the radial displacement of any  mode should behave as
\be\label{eq:XNrs}
\hat X_n(r) =r^{-2} \hat \xi_n(r) e^\phi \propto r\,,
\ee 
which follows from Eq.~\eqref{eq:xidef} and \eqref{eq:correct}. Since in the numerical procedure we impose Eq.~\eqref{eq:bound01}, this  is  a test that we correctly implemented this condition in the discretization  method. In other words  that adding the block matrix in Eq.~\eqref{eq:topleftA} to the top-left corner of the  matrix $A$ does provide the correct behavior close to the stellar center. On the other hand, from Eq.~\eqref{eq:expansionR} we have that
 \be\label{eq:XpR}
X'_n  = a + \omega_n^2 R^3/M e^{- 2\phi}X_n \qquad \text{at} \quad r = R \,, 
\ee
where $a$ is a constant that depends on the background configuration whose explicit expression can be obtained from Eq.~\eqref{eq:expansionR}. The relevant aspect is that $a$     does not depend on $n$. 
Note that in the numerical procedure we have not imposed this condition, indeed we have  discretized close to the right boundary   using the discretized left derivatives in Eq.~\eqref{eq:rightderivatives}. 

From the  plot  reported in Fig.~\ref{fig:radial_1} one can qualitatively see that  the correct linear behavior is reproduced close to the stellar center, as in  Eq.~\eqref{eq:XNrs}, and that close to the stellar surface the derivative of  the radial displacement increases with increasing $n$, as in Eq.~\eqref{eq:XpR}, indeed we recall that $\omega_n<\omega_{n+1}$ and we fixed $\xi_n(R)=1$ for any mode.  More precisely, we find that   the linear behavior close to the stellar center as well as the relation  in Eq.~\eqref{eq:expansionR} close to the stellar center are satisfied with  great accuracy, for instance with the piecewise polytropic we have  an error  less than $\sim 0.1 \%$ already with $N=500$ discretized steps. Note that close to the stellar surface the radial displacement steeply increase: this is due to the fact that  this is region corresponds to the crust, that is light as compared to the stellar interior. 

\begin{figure}[h!]
\includegraphics[width=0.45\textwidth]{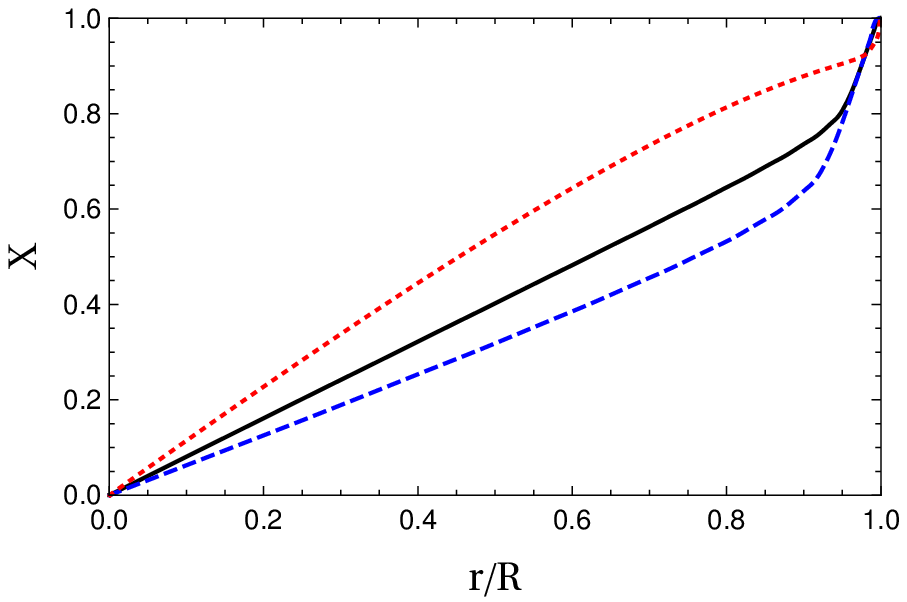}
\includegraphics[width=0.45\textwidth]{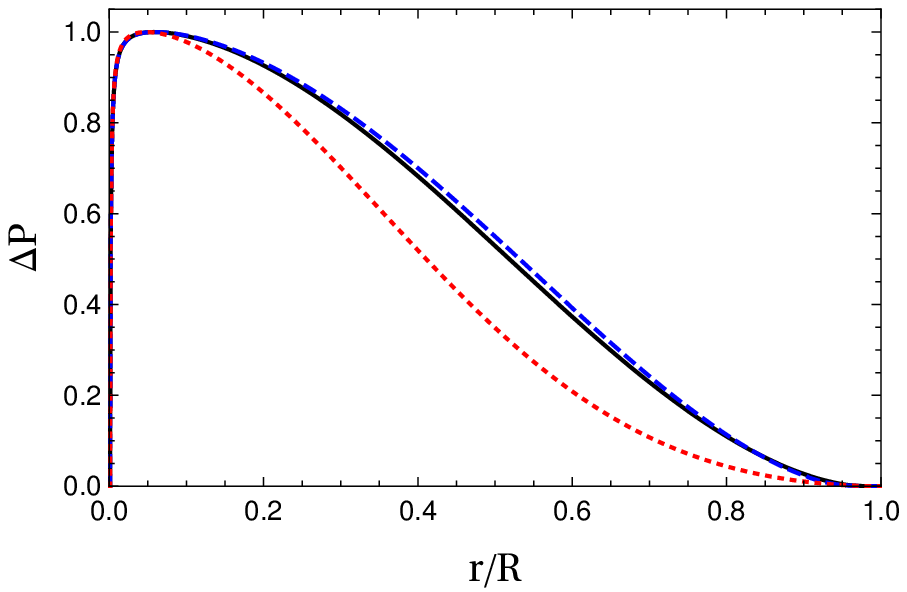}
\caption{Displacement of the fundamental mode (top) and  corresponding Lagrangian fluctuations of the pressure (bottom) obtained with  the SLy4 EoS. The dashed line corresponds to the case $\rho_c=0.73 \times 10^{15}$ g/cm$^3$, mass $M\simeq 1 M_\odot$ and radius $R \simeq 11.3$ km, the solid line corresponds to $\rho_c=0.98 \times 10^{15}$ g/cm$^3$, mass $M\simeq 1.4 M_\odot$ and radius $R \simeq 11.4$ km and the dotted red line to the last stable configuration with $\rho_c=2.846 \times 10^{15}$ g/cm$^3$, mass $M\simeq 2.05 M_\odot$ and radius $R \simeq 9.9$ km.  The numerical algorithm correctly reproduces the linear behavior close to the stellar center.  The pressure oscillations is normalized to the value at the peak.}
\label{fig:Displacement_SLY}
\end{figure}

In Fig.~\ref{fig:Displacement_SLY} we show how  the displacement  of the $0$th radial mode (top panel) and the corresponding pressure oscillations (bottom panel)  changes for the stellar configurations obtained by  the SLy4 EoS at different central densities. In particular, we show the results obtained with  three different values of the central density corresponding to stellar configurations with mass $M\simeq 1 M_\odot $ (solid line), $M\simeq 1.4 M_\odot $ (dashed line)  and to the maximum mass $M\simeq 2.05 M_\odot $ (dotted line). The extended Numerov method correctly reproduces the linear behavior close to the boundaries.  For small central densities, the displacement is peaked at the stellar surface, but with  increasing central density it  tends to become smoother because  more massive stars have a smaller crust. 
The profile of the  Lagrangian oscillations of the pressure shown in the bottom panel of  Fig.~\ref{fig:Displacement_SLY}      are obtained by  Eq.~\eqref{eq:DeltaP}. They reach an extremely small value   at the stellar center, in agreement with  Eq.~\eqref{eq:DeltaPc}, then they exponentially increase with $r$, due to the $e^{\phi}$ term in Eq.~\eqref{eq:DeltaP}, reaching a maximum at a short radial distance from the stellar center.  The position of the peak is almost insensitive to the stellar configuration considered:  the maximum pressure  is always located close to the stellar center. The only difference is that with increasing central density the peak becomes slightly narrower. In all the considered case the pressure oscillation vanishes at the stellar surface. Unfortunately, it does not seem to be possible to infer from this plot that the red dotted line corresponds to the last stable configuration.

\subsection{Piecewise polytropic and hybrid equations of state}

\begin{figure}[h!]
\includegraphics[width=0.45\textwidth]{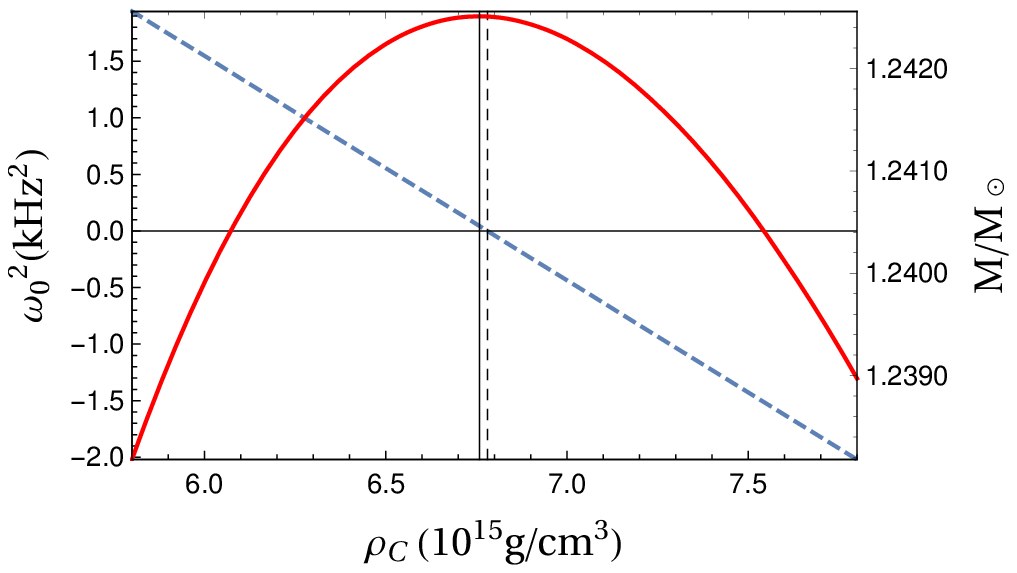}
\includegraphics[width=0.45\textwidth]{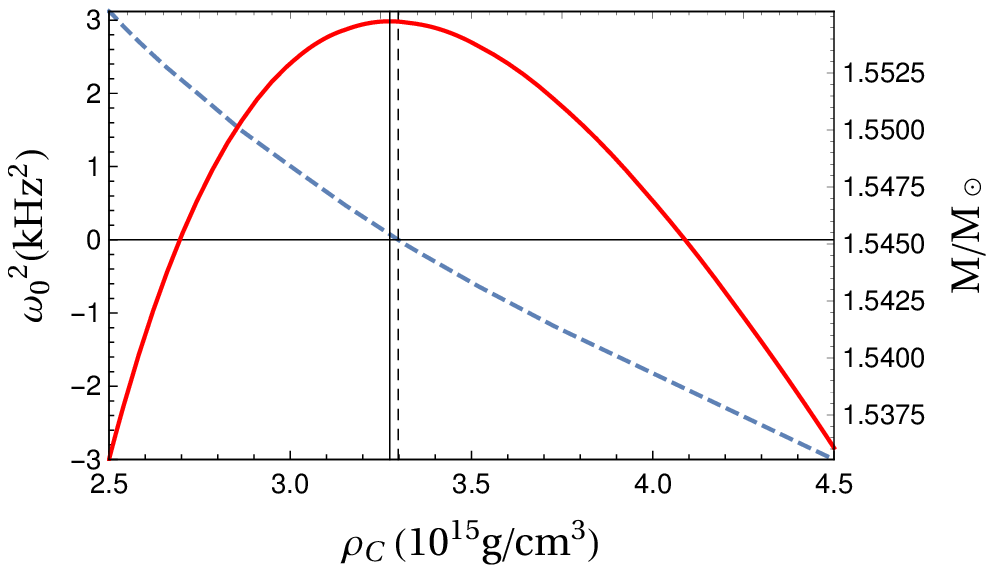}
\caption{Mass and fundamental frequency as a function of the stellar central density for the   piecewise polytope   of Eq.~\eqref{eq:piecepoly}, top panel, and the  hybrid 
EoS in Eq.~\eqref{eq:hybrid}, bottom panel. The solid line corresponds to the mass, in units of the solar mass, the dashed line corresponds to the  frequency squared of the fundamental mode. In both cases the null mode appears at a central density exceeding the value for which  the gravitational mass reaches the maximum value, therefore the last stable configuration is not the one with the maximum mass. The vertical  dashed line indicates to the  null mode  while the solid vertical line indicates  the maximum stellar mass.}
\label{fig:Omega_eos12}
\end{figure}

In Fig.~\ref{fig:Omega_eos12} we report the mass and the fundamental eigenfrequency as a function of the central density for the the piecewise polytropic and  the hybrid EoSs discussed in  Sec.~\ref{sec:background}.  The first has a speed of sound discontinuity, while the second has a speed of sound as well as a matter density discontinuity. Although it is a small effect, we find that in both  cases the last stable configurations, corresponding to the null mode,  have a central density exceeding the one corresponding to  the maximum mass, meaning that there may exist twin stellar configurations with the same gravitational mass but different radii. The results shown for the piecewise polytropic have been obtained with $\Gamma_1=3$ and $\Gamma_2=2$, but swapping the two values we do not find any appreciable increase of the central energy differences. 

\begin{figure}[h!]
\includegraphics[width=0.45\textwidth]{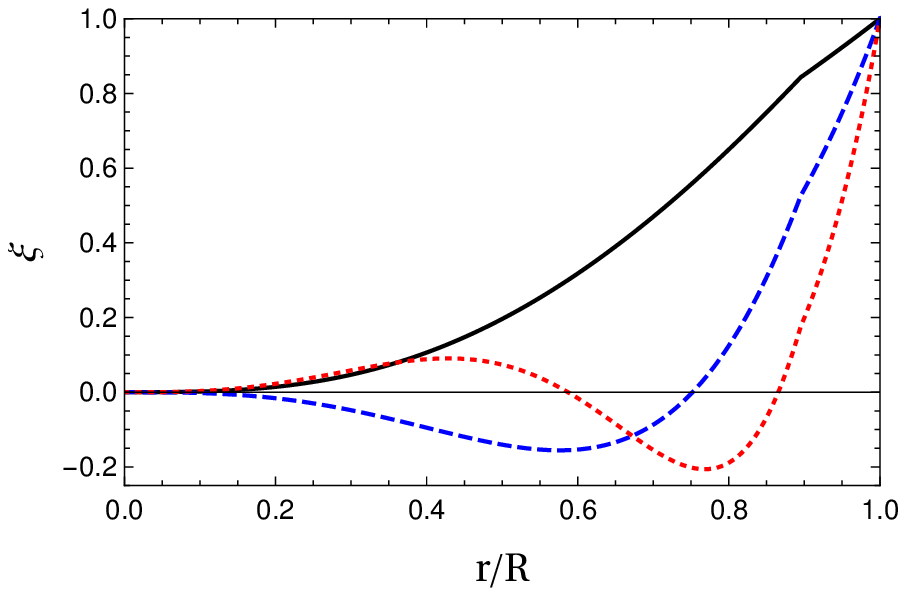}
\includegraphics[width=0.45\textwidth]{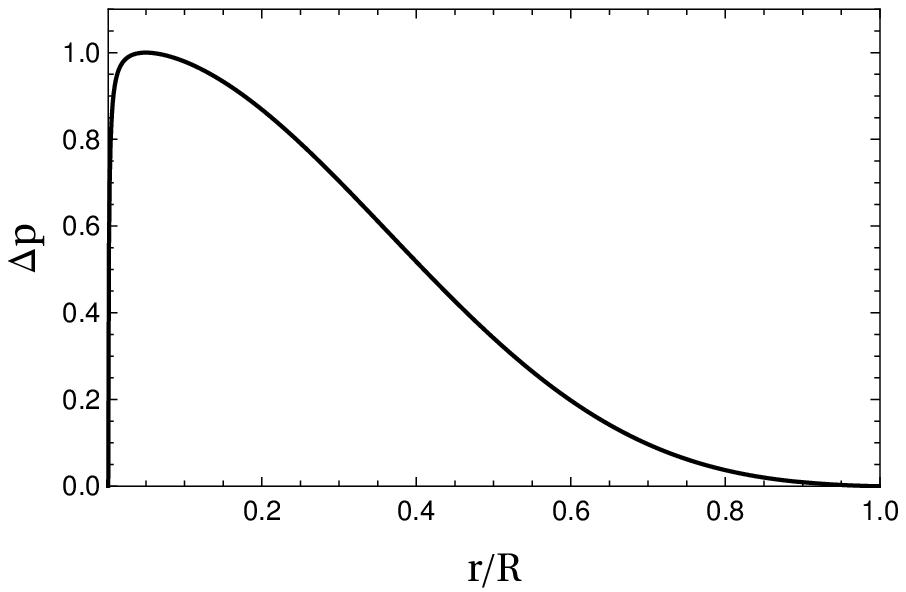}
\caption{Radial profile of the displacement of the fundamental mode (top panel) and of the Lagrangian pressure fluctuation induced by the fundamental mode (bottom panel) for the piecewise polytropic EoS, see Eq.~\eqref{eq:piecepoly}, for $\rho_c =6.5 \times 10^{15}$ g cm$^{-3}$. The displacement has a small kink at  the interface where the speed of sound is discontinuous. The pressure oscillations is normalized to the value at the peak,  it  is  continuous and differentiable at any point. }
\label{fig:Displacement_poly}
\end{figure}

The profiles  of the radial displacement and of the   Lagrangian pressure oscillation obtained with the piecewise polytrope of Eq.~\eqref{eq:piecepoly} are shown in Fig.~\ref{fig:Displacement_poly}. In this case the derivative of the radial displacement is discontinuous at the interface where the speed of sound is discontinuous, which is in agreement with Eq.~\eqref{eq:left_right}. Notice that the interior solution tends to grow more steeply then the external one, which is due to the fact that the derivatives of the displacement at the interface are related by Eq.~\eqref{eq:left_right} and the speed of sound on the left of the interface is smaller than the speed of sound on the right of the interface. Basically,  at the interface   the interior radial displacement   bends to cope with the rapid crust displacement.  We checked  that the kink of $\xi'$ at the interface agrees with Eq.~\eqref{eq:continuity} with great accuracy (the error is at the level of the used numerical accuracy). Notice that  the pressure oscillation shown in the bottom panel of Fig.~\ref{fig:Displacement_poly}  is  continuous and differentiable at any point. It is indeed very similar to the one obtained with the microscopic EoSs discussed above. 

\begin{figure}[h!]
\includegraphics[width=0.45\textwidth]{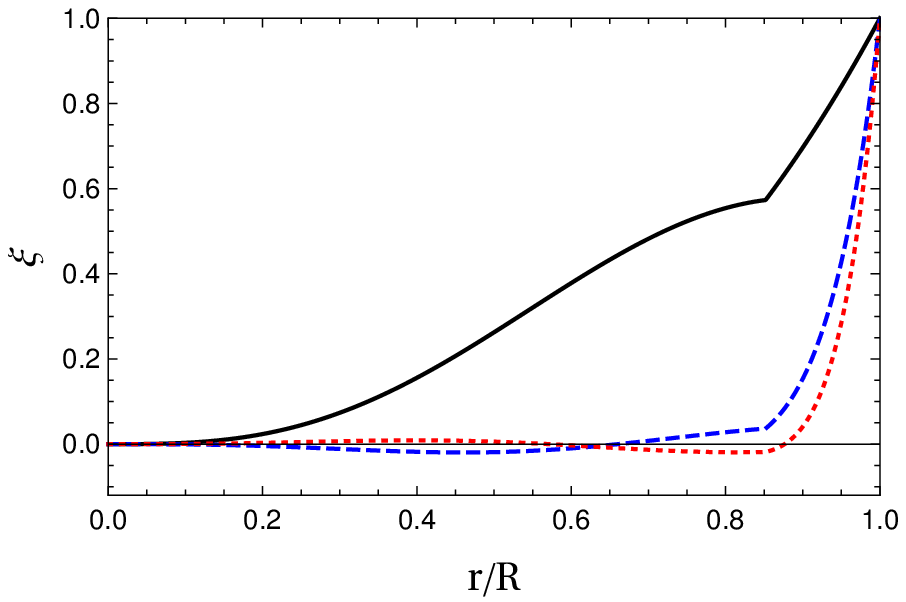}
\includegraphics[width=0.45\textwidth]{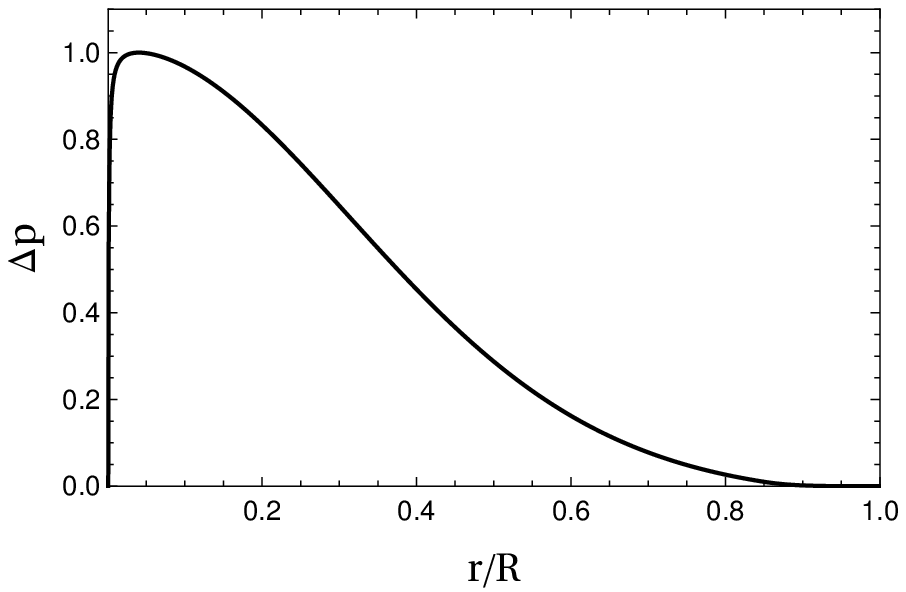}
\caption{Radial profile of the displacement of the first three modes (top panel) and of  Lagrangian pressure fluctuation induced by the fundamental mode (bottom panel)  for the hybrid EoS, see Eq.~\eqref{eq:hybrid}, for $c_s^2=1/3$ and $\rho_c =3.25 \times 10^{15}$ g cm$^{-3}$. The  kink of $\xi$    corresponds to the interface where the matter density is discontinuous. The pressure oscillations is normalized to the value at the peak. The pressure is continuous but, although not visible in the plot, has a small kink  at the interface. }
\label{fig:Displacement_hybrid}
\end{figure}


Then we turn to the  stellar model described by the hybrid  EoS in Eq.~\eqref{eq:hybrid}. We show in Fig.~\ref{fig:Displacement_hybrid} the displacement and the pressure oscillation. Both are continuous and have a kink at the interface between the core and the envelope, but the pressure kink is extremely small and not visible. Note that the displacement of the interior solution tends to become flat, as for self-bound objects, see the discussion after Eq.~\eqref{eq:expansionR}.  As in the previous case at the interface the displacement bends to cope with the  crust displacement. 

\begin{figure}[t!]
\includegraphics[width=0.45\textwidth]{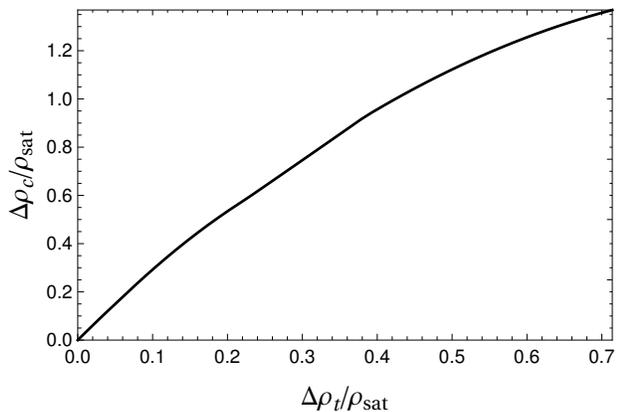}
\caption{Difference between the central energy densities of the last stable configuration and of the maximum stellar mass, Eq.~\eqref{eq:droc},  as a function of the energy density discontinuity at the interface between strange quark matter and the external envelope, see Eq.~\eqref{eq:deltarhot}.}
\label{fig:deltarho}
\end{figure}

In our simple model we can tune the interface energy density jump $\Delta\rho_t$ defined in Eq.~\eqref{eq:deltarhot} to emphasize the effect.  By changing $\rho_1$, which is the largest possible density of the envelope, we can explore  how large  the   energy density difference 
\be\label{eq:droc}
\Delta \rho_c = \rho_{c0} -\rho_{cM} \,,
\ee
can be, where  $ \rho_{cM} $ is central energy density corresponding to  the maximum gravitational mass and $\rho_{c0}$ is the central energy density corresponding to the appearance of the null mode. In Fig.~\ref{fig:deltarho} we report the plot of $\Delta \rho_c$ as a function of 
 $\Delta \rho_t$ for the hybrid stellar model defined in  Eq.~\eqref{eq:hybrid}  with $c_{sq} \simeq 0.28$ and $\Gamma=4/3$. The small value of the speed of sound has been chosen to further emphasize the effect of the interface energy density jump. The resulting maximum mass is about $0.5 M_\odot$, which is indeed small, due to the fact that a small speed of sound implies a large compressibility.  If one considers different values of $c_{sq} $ or $\Gamma$ one obtains similar results, but the effect is even less visible. 

\begin{figure}[h!]
\includegraphics[width=0.45\textwidth]{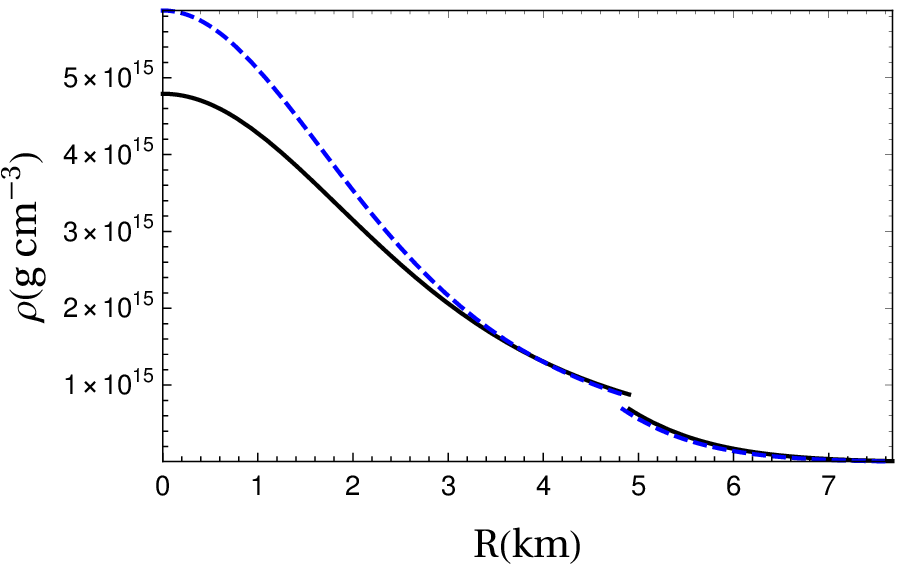}
\includegraphics[width=0.45\textwidth]{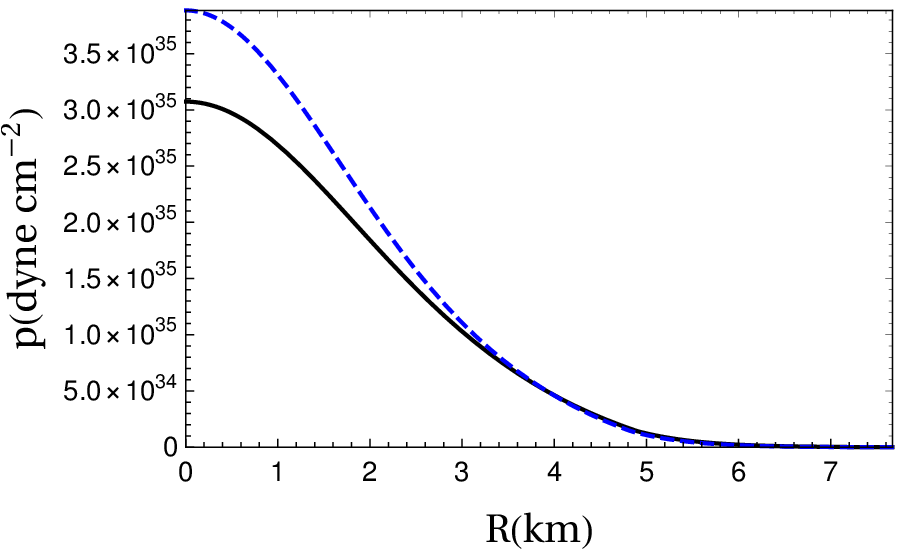}
\caption{Radial profile of the energy density (top panel) and of the pressure (lower panel) for  twin hybrid stars having the same gravitational mass $M\simeq 0.5 M_\odot $ but different central densities: The dashed lines have been obtained considering central densities  $\rho_c \simeq 5.9\times 10^{15}$ g cm$^{-3}$
while for the solid lines $\rho_c \simeq 4.8 \times10^{15}$ g cm$^{-3}$. The stellar 
radii are  $R \simeq 7.43$ Km, dashed line, and $R \simeq 7.67$ Km, solid line.}
\label{fig:density_disc}
\end{figure}

We report in Fig.~\ref{fig:density_disc} the radial profile  of the energy density and of the pressure for twin hybrid  stars having the same gravitational mass $M\simeq 0.5 M_\odot $ and   radii $R \simeq 7.43$ Km (dashed line) and $R \simeq 7.67$ Km solid line.
From this figure it is clear that these twin stars  have a similar envelope, but  the more compact one  accommodates more  hadronic matter in the stellar interior.  We emphasize that both twins are hybrid stars, but the more compact one has more strange matter than the other. We tried several different hybrid star configurations, finding that in any case the radius difference between the twin stars is of the order of hundreds of meters, therefore it is hardly observable.

\section{Conclusions}
\label{sec:conclusions}
We have developed an algorithm to quickly determine the eigenfrequencies and the eigenmodes of the stellar radial oscillations by discretizing the pertinent Sturm-Liouville differential equation. Our method is an extension of the Numerov's method that takes into account the boundary conditions and the possible discontinuous interfaces.
We find that the method is fast and precise for any considered EoS. Indeed, it gives radial displacements and  pressure fluctuations that are smooth functions of the radial coordinate and that are in agreement with the foreseen behavior at the boundaries. Moreover, it allows us to reliably describe the  interfaces between different states  of matter.  

An important aspect is that  the extended Numerov method efficiently works for many different  stellar model, as an example we considered three microscopic EoSs, a  joined polytrope and an  hybrid stellar models.  In any considered case we find that the algorithm is very fast and the results are extremely stable for $N=1000$ discretized points. Taking  $N=1500$ points we do not  find any appreciable change in the eigenfrequencies or in the eigenmodes.   Then, the  diagonalization of the $N\times N$ matrix $A$ in Eq.~\eqref{eq:eigeneq} only requires few seconds on a laptop computer. The remarkable point is that in this way  one obtains $N$ eigenfrequencies and eigenmodes, meaning that one can model with great accuracy any stellar radial oscillation  by a Fourier decomposition.

In the presence of an interface discontinuity, the algorithm isolates the singular point and properly expands the displacement on the left and on the right of the discontinuity. We find that with  a speed of sound or with a matter density discontinuity  the last stable and the maximum mass configurations do not coincide.  This allows the existence of twin compact stars, that is  stars with the same mass but different radii. Therefore, we confirm the results of~\cite{VasquezFlores:2012vf, Pereira:2017rmp} for hybrid stars and extend it to any piecewise polytropic solutions, even in the presence of only a speed of sound discontinuity.  For hybrid stars we have tuned the density discontinuity to study how large  can be the difference between the two critical densities finding that  this difference tends to grow as depicted in Fig.~\ref{fig:deltarho}.  In any considered case, the radius  difference between twin partners is small,  of few hundred meters, at most, therefore  they can be hardly discriminated by  observation, if they exist.

In the present work we focused on the linear response analysis, however the nonlinear effects may qualitatively change the picture~\cite{1982AcA....32..147D, Gabler:2009yt}, with  a much richer dynamics. It would be interesting to analyze the behavior of nonlinear effects in the presence of interface discontinuities.

\acknowledgments
We thank Ignazio Bombaci and Fabrizio Nesti for several discussions during the preparation of this work.

%
%
%

\appendix
\section{Damping}
\label{app:damping}
In the following we  focus on the discontinuity in $c_s$ keeping $\rho$ continuous, but the procedure can be  extended to the case with both discontinuous $c_s$ and $\rho$. Since $W, Q$ and $\xi$ are continuous function, we have from Eq.~\eqref{eq:sturm-liouville} that 
\be\label{eq:interface}
(H \xi')'_L \simeq (H \xi')'_R\,,
\ee
and we notice that $H \xi' = - \Delta P e^{\lambda + 2 \phi}$, therefore we can rewrite the above equation as
\be
(\Delta P_L)' +\Delta P_L (\lambda' + 2 \phi')_L \simeq (\Delta P_R)' +\Delta P_R (\lambda' + 2 \phi')_R\,,
\ee
where  both $\phi'$ and $\lambda'$ are continuous functions, see Eqs.~\eqref{eq:csdisc}. It follows that if $\Delta P$ is continuous then also $\Delta P'$ is continuous (this will not be the case when $\rho$ is discontinuous). Therefore, even in the presence of a speed of sound discontinuity $\Delta P$ is  a smooth function of $r$.

The differential equation at the boundary allows us to explore the possible effect of wave damping in a certain region of the star. If we define $\alpha = c_s^2 \xi'$, then
the interface condition in Eq.~\eqref{eq:interface} can be written as
\be\label{eq:alpha}
\alpha'_L + K_L \alpha_L = \alpha'_R + K_R \alpha_R 
\ee
where
\be
K = (B+ E)/D\, 
\ee
and 
\begin{align}
B & = \log(r^-2 e^{\lambda + 3 \phi})'\\
D & = e^{2(\phi -\lambda)}\\
E & = D \frac{p' + w'}{p+\rho}\,. 
\end{align}

The  continuity of the pressure implies that $\alpha$ is a continuous function, but $K$ is  discontinuous and this means   that the relation between the second order derivatives of the displacement at the interface  is nontrivial. However,
in the special cases in which one phase is characterized by a large bulk viscosity,  the pressure perturbation vanishes at the interface and $\xi$ is stationary (it has a maximum or a minimum) and then $\alpha_L= \alpha_R=0$.   From Eq.~\eqref{eq:alpha} it follows that $\alpha'_L =\alpha'_R$. This special boundary condition is therefore 
\begin{align}
\xi'_L &=\xi'_R=0 \\
c_L^2 \xi''_L &=c_R^2 \xi''_R\,.
\end{align}
If we further assume that  $\xi''_L = \xi''_R =0$, we obtain   that 
\be
\xi(\bar r)=0\,.
\ee
Since the fundamental mode cannot have a node, it follows that only in this particular case one can really separate the modes as core modes and surface modes~\cite{Gondek:1999ad}, depending on whether the internal or external displacement is nonvanishing.  
%

\bibliography{BIB}
\bibstyle{comp}

\end{document}